\documentclass[%
 aip,
 amsmath,amssymb,
 reprint,%
]{revtex4-2}

\usepackage{graphicx}
\usepackage{bm}
\graphicspath{{figures/}}
\draft 
\usepackage{amsmath}
\usepackage{siunitx}
\usepackage{physics}
\usepackage{color}
\usepackage[makeroom]{cancel}

\newcommand{\eqr}[1]{Eq.\thinspace(\ref{eq:#1})}

\newcommand{\pfrac}[2]{\frac{\partial #1}{\partial #2}}
\newcommand{\fgr}[1]{Fig.\thinspace\ref{fig:#1}}

\newcommand{\gke}{\texttt{Gkeyll}}

\begin{document}


\title{An investigation of shock formation versus shock mitigation of colliding plasma jets}       



\author{Petr Cagas}
\email[]{pcagas@vt.edu}
\affiliation{Kevin T. Crofton Department of Aerospace and Ocean
  Engineering, Virginia Tech, Blacksburg, VA, 24061, USA}

\author{James Juno}
\affiliation{Princeton Plasma Physics Laboratory, Princeton, NJ,
  08540, USA}

\author{Ammar Hakim}
\affiliation{Princeton Plasma Physics Laboratory, Princeton, NJ,
  08540, USA}

\author{Andrew LaJoie}
\affiliation{Los Alamos National Laboratory, Los Alamos, NM, 87544, USA}

\author{Feng Chu}
\affiliation{Los Alamos National Laboratory, Los Alamos, NM, 87544, USA}

\author{Samuel Langendorf}
\affiliation{Los Alamos National Laboratory, Los Alamos, NM, 87544, USA}

\author{Bhuvana Srinivasan}
\email[]{srinbhu@vt.edu}
\affiliation{Kevin T. Crofton Department of Aerospace and Ocean
  Engineering, Virginia Tech, Blacksburg, VA, 24061, USA}


\date{\today}

\begin{abstract}
This work studies the interaction between colliding plasma jets to
understand regimes in which jet merging results in shock formation
versus regimes in which the shock formation is mitigated due to the
collisionless interpenetration of the jets.  A kinetic model is
required for this study because fluid models will always produce a
shock upon the collision of plasma jets. The continuum-kinetic,
Vlasov-Maxwell-Dougherty model with one velocity dimension is used to
accurately capture shock heating, along with a novel coupling with a
moment equation to evolve perpendicular temperature for computational
efficiency.  As a result, this relatively inexpensive simulation can
be used for detailed scans of the parameter space towards predictions
of shocked versus shock-mitigated regimes, which is of interest for
several fusion concepts such as plasma-jet-driven magneto-inertial
fusion (PJMIF), high-energy-density plasmas, astrophysical phenomena,
and other laboratory plasmas.  The initial results obtained using this
approach are in agreement with the preliminary outcomes of the Plasma
Liner Experiment (PLX).
\end{abstract}


\maketitle 


\section{Introduction}

The study of collisionless and collisional shock formation
\cite{biskamp1973collisionless, liberman1986physics} has remained an
open area of research in a wide variety of plasmas ranging from the
laboratory to astrophysics. An understanding of plasma shocks can have
significant implications for a range of fusion concepts that include
inertial confinement fusion \cite{casanova1991kinetic} and
plasma-jet-driven magneto-inertial fusion (PJMIF),
\cite{Thio1999magnetized,Hsu2012spherically,Knapp2014possible} where
shocks can be detrimental to the uniformity of liner formation during
implosion.  The relevance of colliding plasma jets goes beyond
fusion-relevant experiments.  A number of astrophysical phenomena
require an understanding of plasma jets, particularly at high Mach
numbers, where the jets exhibit strong internal collisions but do not
collide with each other.\cite{Gregory2009colliding,Ryutov2012intra}
Additionally, collisional jets play a key role in several laboratory
basic plasma science \cite{mohammed2022ion} and warm dense matter
(WDM) experiments.\cite{Falk2018experimental}

This work studies the regimes in which shocks form upon collisions
between plasma jets compared to regimes where the shocks are mitigated
and the jets interpenetrate with minimal interaction.  In all of the
regimes explored here, each plasma jet is sufficiently collisional
that it self-thermalizes and propagates in local thermodynamic
equilibrium.  However, the interaction between plasma jets is not
necessarily dominated by collisional interaction and a careful
examination of the parameter space is necessary to determine shock
formation versus shock mitigation upon jet merging.

While this study is generally applicable across a range of plasmas,
the plasma jets considered in this work are chosen based on parameters
for the PJMIF experiments. These jets typically have a high enough
Mach number that jets may form even at oblique angles. The resulting
shock heating decreases the Mach number and the peak stagnating
pressure, which can present challenges with maintaining spherical
symmetry and uniformity during an implosion.\cite{Hsu2012spherically}
The formation of such shocks has already been confirmed both
experimentally and with numerical simulations for multiple jets with
varied angles between each pair of the guns.\cite{Case2013merging,
  Merritt2014experimental}

It is worth pointing out, that fluid models, which evolve macroscopic
quantities like density, momentum, and energy, intrinsically assume
that particles of each species always have the Maxwellian
distribution.  Therefore, a fluid model will always predict a
collisional interaction between impeding plasma jets and thus produce
a shock, when in reality jet merging is not guaranteed to produce a
shock if the interaction between the jets is sufficiently
collisionless.  Depending on the parameter regime, there could be
complete interpenetration of the jets. To accurately capture the
physics of merging plasma jets, a kinetic model is required.  In
addition, as we will show, accurate jet merging studies require a
kinetic model which correctly captures all the degrees of freedom of
the plasma.

Here, continuum kinetic simulations of merging jets are performed with
normal incidence using the \gke{} plasma simulation framework
(\url{https://gkeyll.readthedocs.io}).  A predictive capability is
developed to understand regimes of shock formation versus shock
mitigation in the parameter space of the Plasma Liner Experiment
(PLX).\cite{Hsu2012spherically, Hsu2015laboratory, Hsu2017experiment}
Doppler shifts of the ArII \SI{434.8}{nm} emission on PLX plasma jets,
measured on a high-resolution McPherson 2062 Scanning Monochromator
with a 2400-\si{mm^{-1}} grating and \SI{4}{m} focal length in the
double-pass configuration as used here, provide an insight into the
ion distribution function which is used to validate the distribution
functions obtained from \gke{}.

This work first provides a description of the kinetic numerical model
used in \gke{}, emphasizing the hybrid reduction to one velocity space
(Section\,\ref{sec:model}). Next, Section\,\ref{sec:sims} focuses on
the simulation setup and the assessment of both the collisional model
and the hybrid reduction of the velocity space. Finally,
Section\,\ref{sec:scaling} shows effects of scaling relevant
parameters and the comparison to PLX's experimental data are presented
Section\,\ref{sec:plx}.

\section{Numerical model}\label{sec:model}

All the cases presented here are simulated using the
Vlasov-Maxwell-Dougherty\cite{Juno2018discontinuous,
  Hakim2020conservative, Hakim2020alias} model in the \gke{}
simulation framework (see a note at the end of this paper for details
on how to install \gke{} and obtain the input files);
plasma species, $s$, are evolved individually using
  \begin{equation}\label{eq:vlasov}
    \pfrac{f_s}{t} + \bm{v}\cdot\nabla_{\bm{x}}f_s +
    \frac{q_s}{m_s}\left(\bm{E} +
    \bm{v}\times\bm{B}\right)\cdot\nabla_{\bm{v}}f_s = \left(\frac{\dd{f_s}}{\dd{t}}\right)_c,
  \end{equation}
  where $f$ is the particle distribution function, $q$ and $m$ are
  particle charge and mass, respectively. Electromagnetic fields,
  $\bm{E}$ and $\bm{B}$, and collisions,
  $\left(\dd{f_s}/\dd{t}\right)_c$ couple the species together.  The
  fields are evolved using the Maxwell's equations, which in 1D cases
  shown in this work reduce to the Ampere's law,
  \begin{equation}
    \frac{\dd{E}}{\dd{t}} = -\frac{J}{\varepsilon_0},
  \end{equation}
  where the current, $J$, is calculated by combining moments of the
  distribution functions.

  Collisions are applied using the reduced Fokker-Planck
  operator\cite{Hakim2020conservative,Francisquez2022improved} (FPO;
  the reduced version is often referred to as Dougherty or
  Lenard-Bernstein operator),
  \begin{equation}\label{eq:colld}
    \left(\frac{\dd{f_s}}{\dd{t}}\right)_c = \sum_{r}\nu_{sr}\nabla_{\bm{v}}\cdot\left(\left(\bm{v}-\bm{u_{sr}}\right)f_s +
    \frac{T_{sr}}{m}\nabla_{\bm{v}} f_s\right),
  \end{equation}
  where the summation over species accounts for inter-species
  collisions.  $\bm{u_{sr}}$ and $T_{sr}$ are bulk velocity and temperature,
  which are calculated from the moments of the distribution function,
  $f_s$. For intra-species collisions, the moments are calculated as 
  \begin{align}\label{eq:moments1}
    M_0 &= \int f \dd{v_x} = n,\\
    M_1 &= \int v_xf \dd{v_x} = n\mathbf{u},\label{eq:moments2}\\
    M_2 &= \int v_x^2 f \dd{v_x} = n \mathbf{u}^2 + d n
    \frac{T}{m},\label{eq:moments3}
  \end{align}
  where $d$, is the number of velocity dimensions. Correction for
  finite velocity space extends is also required for energy
  conservation.\cite{Hakim2020conservative} These moments are also
  used to construct additional moments for inter-species
  interaction.\cite{Francisquez2022improved} Inter-species collisions
  are included in this. $\nu_{sr}$ in \eqr{colld} is the collision
  frequency.

The system of equations is then discretized using the discontinuous
Galerkin (DG) method\cite{Cockburn1998runge, Cockburn2001runge,
  Hesthaven2007nodal, Juno2018discontinuous, Hakim2020alias} with the
Serendipity basis.\cite{Arnold2011serendipity}

\eqr{vlasov} is written generally for any number of spatial and
velocity dimensions. Frequently, plasma dynamics can be captured with
reduced dimensional phase-space, for example, a one-dimensional
electrostatic problem can be modeled with a single velocity dimension
because the collisionless forces on the plasma, the resulting
electrostatic electric field, are dominantly in a single velocity
dimension. While such reduction significantly decreases computational
cost, it is not always sufficient to capture the relevant
physics. This is true, for example, for moderately collisional regimes
where both collisional and collisionless physics is important. To
attain the correct particle distributions in such cases, it is
necessary to use the right amount of the velocity space dimensions.

However, simulations with three velocity space dimension are
significantly more expensive and hence are not well suited for large
parameter scans, which are the motivation of this work. To overcome
this, we chose an alternative approach and incorporated the effect of
three velocity space dimensions using a method described in previous
work.\cite{Cagas2017continuum} It is based on rewriting the full
distribution function, $f(x,\bm{v})$, in such a way that the
perpendicular velocity dimensions are decoupled,
\begin{equation}
  f(x,\bm{v}) = f_x(x,v_x)\underbrace{\frac{m}{2\pi
      T_\perp(x)}\exp\left(-m\frac{v_y^2+v_z^2}{2T_\perp(x)}\right)}_{f_\perp(x,v_y,v_z)}.
\end{equation}
This requires the assumption that the particles retain the Maxwellian
distribution in the perpendicular directions. The non-equilibrium
dynamic part, $f_x(x,v_x)$, is then directly evolved with the Vlasov
equation, \eqr{vlasov}, while the perpendicular component,
$f_\perp(x,v_y,v_z)$, is coupled to the system using an advection
equation (see \eqr{perptemp}).  This implicitly assures the correct
number of velocity dimensions in the system without the need to fully
discretize a higher-dimensional phase-space. Note that since $\iint
f_\perp dv_y dv_z = 1$,
\begin{equation}
\iiint f(x,\bm{v}) \dd{\bm{v}} = \int f_x(x,v_x) \dd{v_x} = n(x).
\end{equation}

The temperature in \eqr{vlasov} includes both the parallel and
perpendicular components,
\begin{equation}
  T=\frac{1}{3}\left(T_x + 2T_\perp\right),
\end{equation}
where $T_x$ is calculated from the moments of $f_x(x,v_x)$.

The equation for perpendicular temperature is derived by taking the
$v_y^2 + v_z^2$ moment of the kinetic equation,
\begin{equation}\label{eq:perptemp}
  \pfrac{nT_\perp}{t} + \pfrac{}{x}\left(u_xnT_\perp\right) =
  n\nu\left(T-T_\perp\right).
\end{equation}

Finally, the conservation properties of the collisional model need to
be addressed. The previous work of Hakim et
al.\cite{Hakim2020conservative} showed that for collisions to conserve
momentum and energy, moments of the distribution function,
Eq.\thinspace(\ref{eq:moments1})\thinspace-\thinspace(\ref{eq:moments2}),
have to be corrected for finite velocity-space extents,
\begin{equation}\label{eq:momc}
  \frac{T}{m}\left[f\left(v_{max}\right)-f\left(v_{min}\right)\right] +
  M_1 - u_xM_0 \doteq 0,
\end{equation}
\begin{multline}\label{eq:enc}
  \frac{T}{m}\left[v_{max}f\left(v_{max}\right)-v_{min}f\left(v_{min}\right)\right]
  +\\ M_2 - u_xM_1 -M_0\frac{T}{m} \doteq 0.
\end{multline}

Here, in the case of the hybrid model with perpendicular temperature,
the correction needs to be modified to
\begin{multline}\label{eq:enc2}
  \frac{T}{m}\left[v_{max}f\left(v_{max}\right)-v_{min}f\left(v_{min}\right)\right]
  +\\ M_2 + 2M_0\frac{T_\perp}{m}- u_xM_1 -3M_0\frac{T}{m} \doteq 0.
\end{multline}
The $\doteq$ symbol
denotes weak equality. We say that two functions, $f$ and $g$,
are weakly equal over an interval $I$ if
\begin{equation}
  \int_I \big(f(x)-g(x)\big) \psi(x) \dd{x} =0,
\end{equation}
where $\psi$ is a basis function. See Ref.
[\onlinecite{Hakim2020conservative}] for more details.

In this work, the distribution function $f_{x}$ is evolved using
1D-simplification of \eqr{vlasov},
  \begin{equation}\label{eq:vlasov1d}
    \pfrac{f_{x,s}}{t} + v_x\pfrac{f_{x,s}}{x} +
    \frac{q_s}{m_s}E_x\pfrac{f_{x,s}}{v_x} =
    \left(\frac{\dd{f_{x,s}}}{\dd{t}}\right)_c,
\end{equation}
and the perpendicular temperature is evolved using \eqr{perptemp}.
This model is referred to as the
Parallel-Kinetic-Perpendicular-Moments (PKPM) model.

\section{Simulation setup and model assessments}\label{sec:sims}

This work focuses only on the interaction of the jets. To conserve
computational time, each simulation is set to start right before the
plasma jets come into contact, i.e., the propagation of the jets is
not simulated.

An example of such initial conditions in phase-space is presented in
\fgr{init} as Maxwellian distribution functions for electrons (a) and
singly-charged argon ions (b). In this case, the initial temperature
is set to \SI{1.5}{eV} for both species and the bulk velocity of each
jet is \SI{15}{km/s}, i.e., the relative velocity between the jets is
\SI{30}{km/s}. Note that both species have the same bulk velocity and
there is no initial electric current.

\begin{figure}
  \centering
  \includegraphics[width=\linewidth]{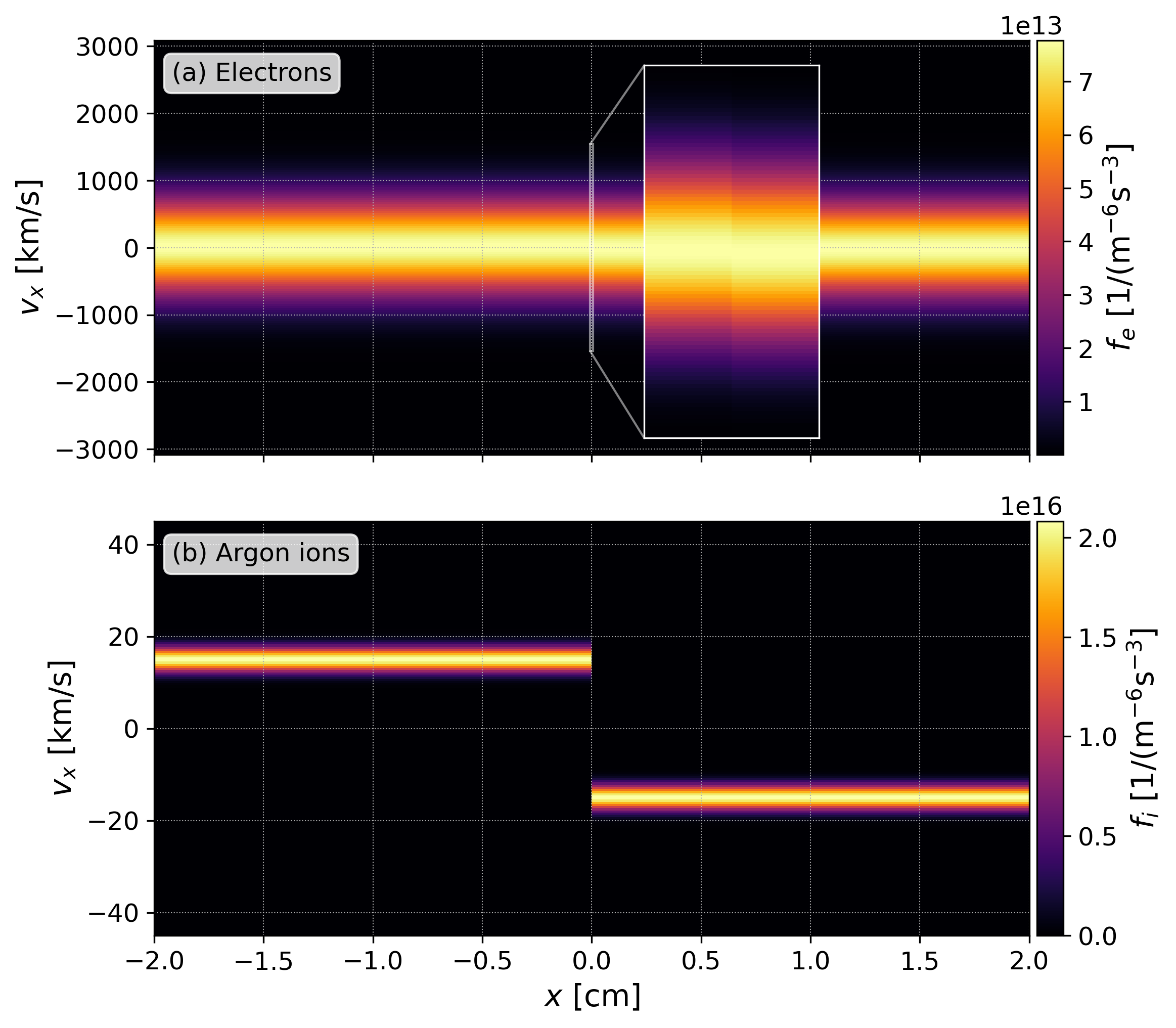}
  \caption{Initial conditions of the electron and ion distribution
    function for two \SI{15}{km/s} argon jets over the whole
    \SI{4}{cm} domain. The initial temperature is set to \SI{1.5}{eV}
    for both species. As the thermal velocity of electrons is, at this
    temperature, high relative to the jet velocity, the initial
    discontinuity in the electron distribution function is barely
    noticeable even within the expanded scale presented.}
    \label{fig:init}
\end{figure}

\subsection{Effectiveness of the PKPM model}

\begin{figure}
  \centering
  \includegraphics[width=\linewidth]{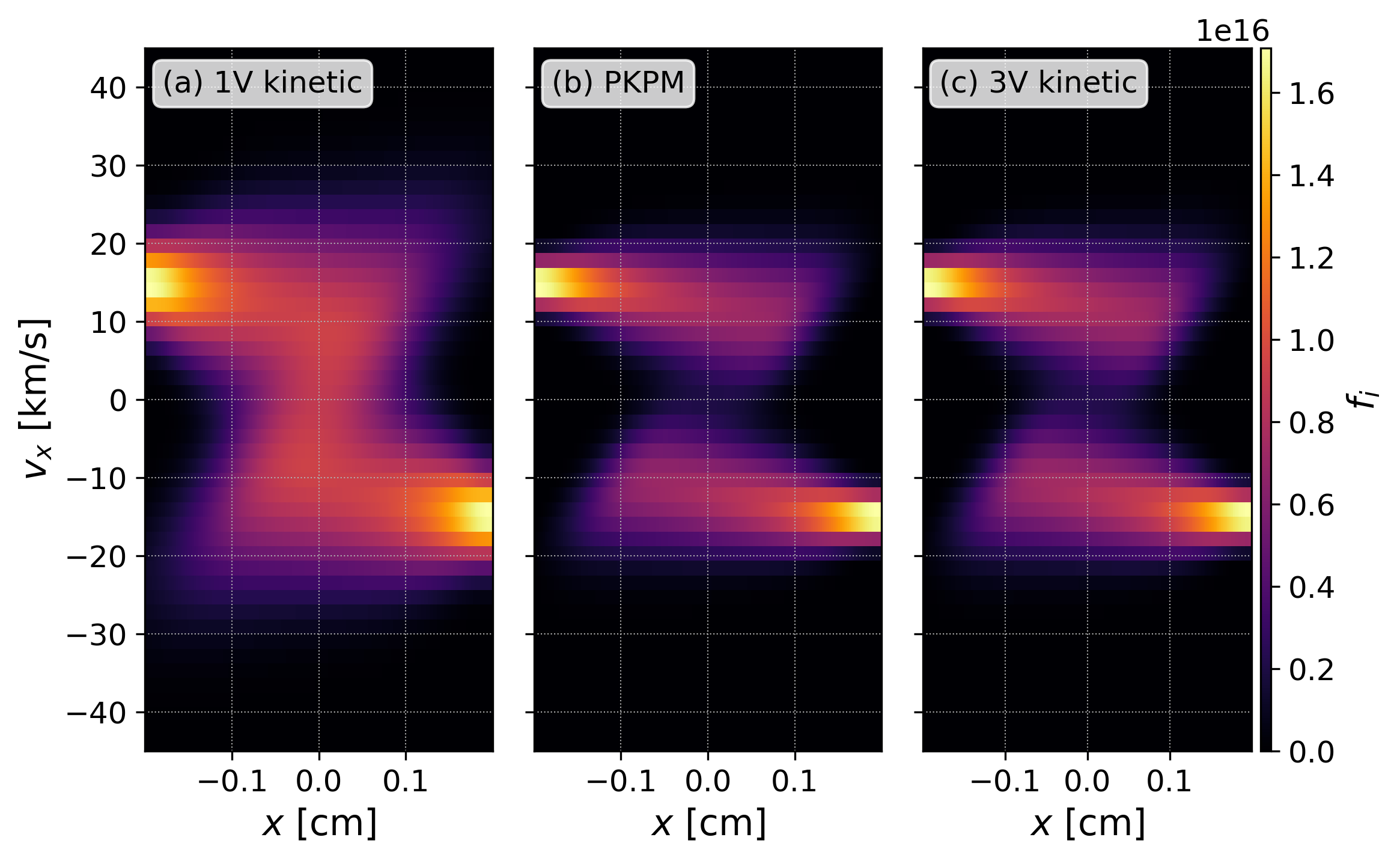}
  \caption{Comparison of simulations with one velocity dimension (1V),
    hybrid simulation with kinetic description in the $x$-direction
    coupled with parallel temperature evolution (PKPM), and with full
    three-dimensional velocity space (3V). The full 3V distribution
    function has been integrated over the $v_y$ and $v_z$ directions
    for comparison. All three cases are evolved for \SI{0.1}{\micro s}
    from the initial contact.  Note that the PKPM and 3V solutions are
    in agreement and they differ from the 1V solution.}
    \label{fig:model_comp1}
\end{figure}

To demonstrate the effectiveness of the hybrid PKPM model, we run
three simulations from the same initial conditions in computational
domains with one (1V) and three (3V) velocity
dimensions. Figure\thinspace\ref{fig:model_comp1} shows the three ion
distribution functions after \SI{0.1}{\micro s} from the initial
contact. All these cases use the same collisional model, the only
differences are the number of velocity dimensions and the extra
perpendicular equations in the PKPM model. Note that the velocity
resolution is slightly reduced in comparison to \fgr{init} to make
this test less computationally expensive. Panel (a) shows the 1V
model, panel (b) the PKPM model, and panel (c) the 3V model. The full
3V distribution function in the latter case has been integrated over
the $v_y$ and $v_z$ directions for
comparison. Figure\thinspace\ref{fig:model_comp1} shows the remarkable
accuracy of the PKPM model as compared to the full 3V simulation.
  
\begin{figure}
  \centering
  \includegraphics[width=\linewidth]{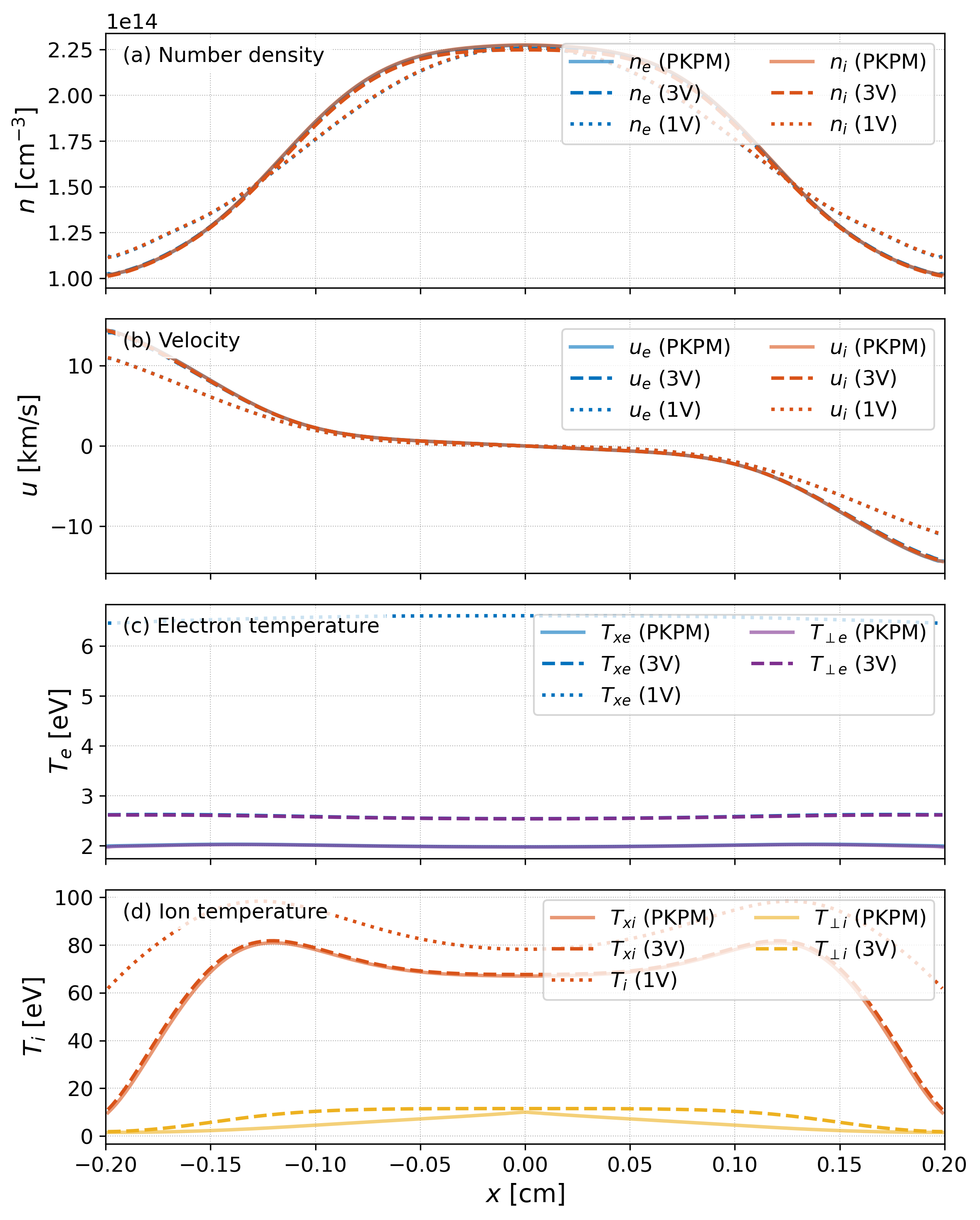}
  \caption{Moments of the distribution functions in \fgr{model_comp1};
    (a) number density, (b) particle flux, (c) electron temperatures,
    and (d) ion temperatures. The goal is to establish how the hybrid
    PKPM model agrees with the 3V model relative to the 1V
    model. Densities, fluxes, and parallel ion temperatures show a
    good agreement. The electron temperatures are underestimated but
    have better agreement than the 1V model, and ion thermalization in
    the perpendicular dimension lags behind the 3V model. Note that in
    this case, the parallel and perpendicular electron profiles are on
    top of each other due to high electron-electron collision
    frequency.}
    \label{fig:model_comp2}
\end{figure}

Figure\thinspace\ref{fig:model_comp2} supplements the distribution
functions from \fgr{model_comp1} with integrated moments; number
density, flux, and temperature. The PKPM and 3V models agree well for
density and flux.  Even though the electron temperature is
underestimated in the PKPM model, it is still more accurate than in
the 1V model. Finally, ion parallel temperatures match while the heat
transfer to the perpendicular direction lags in the PKPM model in
comparison to 3V.

This gives us confidence that PKPM can successfully replicate 3V
simulations for these cases. The PKPM model provides access to
computationally-efficient fully explicit kinetic simulations on a
domain spanning several centimeters while resolving the collisional
mean-free-path for a couple of \si{\mu s}.  Simulations using the PKPM
model take just a few hours on a single GPU. The computational cost is
only marginally higher than for a 1V simulation. These comparisons
demonstrate that the 1V model produces substantially different results
and is not accurate for these problems.

\subsection{Collision of cold beams}

In the \SI{15}{km/s} case (\fgr{init}), both species collide to form a
Maxwellian distribution with zero bulk velocity. The higher collision
frequency of electrons results in a small variance from the Maxwellian
distribution across the whole simulation domain while the heavier
argon ions form non-Maxwellian transition regions (approximately
between 0.5 and \SI{0.75}{cm} in \fgr{evolved}b). Note that the
trapping potential (i.e., the velocity $v_x$ for which $1/2 m v_x^2 =
|q\phi|$), represented by the green line in \fgr{evolved}a encompasses
a trapping region.  In the case of collisionless electrostatic shocks,
electrons typically fill the trapping region forming a ``flat-top''
distribution.\cite{Forslund1970formation, Pusztai2018low,
  Sundstrom2019effect} Here, electrons retain the Maxwellian velocity
distribution while expanding beyond the trapping region due to their
high collision frequency.

\begin{figure}
  \centering
  \includegraphics[width=\linewidth]{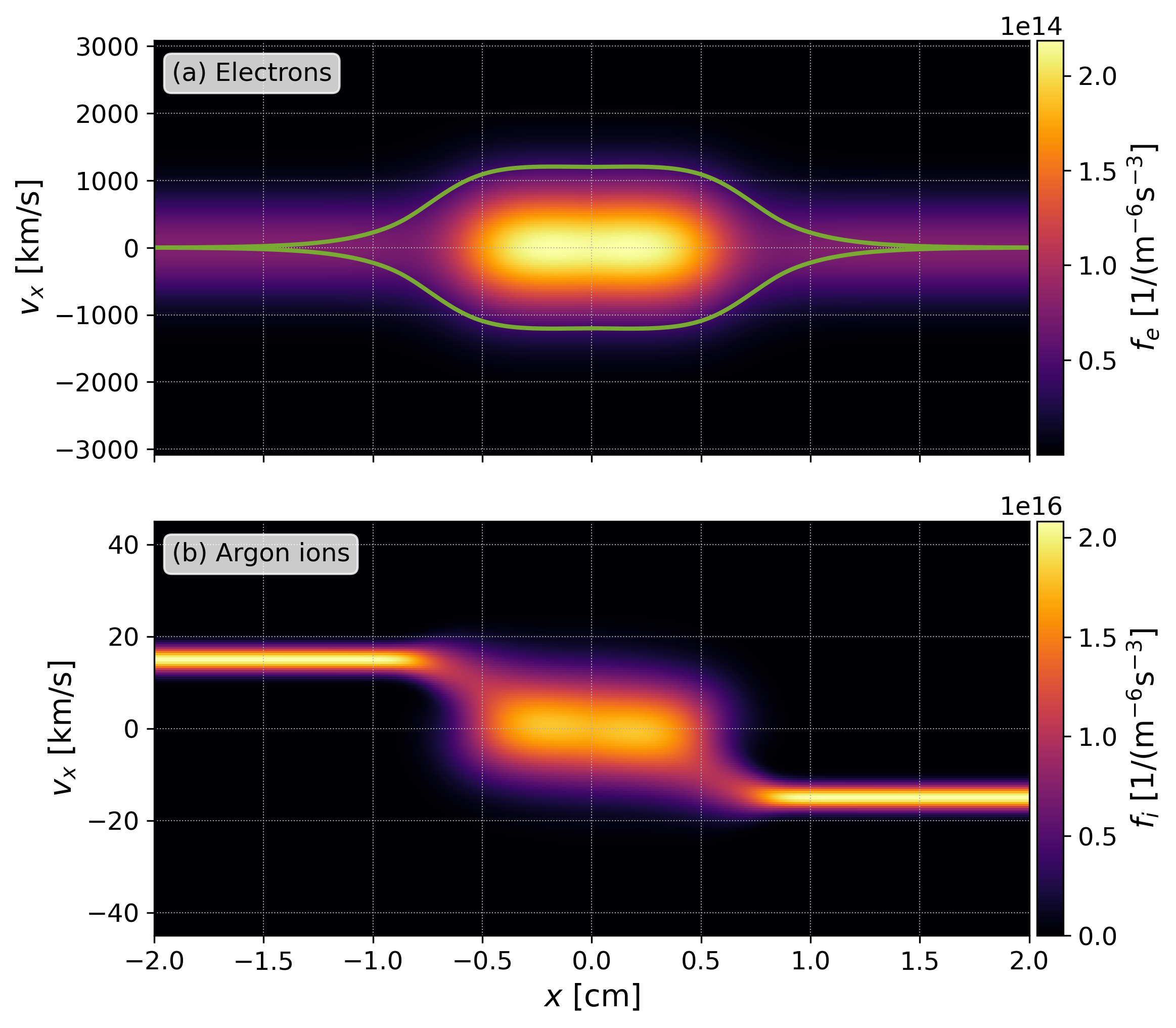}
  \caption{Electron and ion distribution functions, evolved for
    \SI{1}{\micro s} from the initial conditions shown in \fgr{init}
    (the point of contact). The ion populations collide, forming a
    Maxwellian distribution with zero bulk velocity and increased
    density. Additionally, the green line in the electron plot
    corresponds to the trapping potential, i.e., the velocity $v_x$
    for which $1/2 m v_x^2 = |q\phi|$. Note that the potential is
    calculated during postprocessing and the simulation uses
    Vlasov-Maxwell model.}
    \label{fig:evolved}
\end{figure}

This simulation uses the Spitzer formula to calculate
  the collision frequency in \eqr{colld}. As we will show later in
  \fgr{experiment}, these results are in agreement with experimental
  measurements. However, this was not the case for the faster,
  \SI{50}{km/s} \SI{1.5}{eV} jets. There the same simulation approach
  still results in a merge while the experimental results show
  interpenetration. This is due to the fact that the reduced
  collisional model assumes a collisional frequency independent of
  velocity and overestimates the effect of collisions for tails of the
  distribution function. Note that in the case of the faster jets, the
  thermal velocity is significantly lower than the bulk velocity
  ($v_{th}=\SI{1.9}{km/s}$). 

To capture the collisions between two cold beams, the full Fokker-Planck
operator\cite{Rosenbluth1957fokker} is required;
\begin{equation}\label{eq:fpo}
  -\frac{1}{2}\nabla_v \cdot \big( \bm{a}f -
  \bm{D}\cdot\nabla_vf\big),
\end{equation}
where $\bm{a}$ and $\bm{D}$ are calculated from the Rosenbluth
potentials, $h(\bm{x},\bm{v})$ and $g(\bm{x},\bm{v})$.

We can get a better understanding of the interaction between the
beams, by approximating an effective collision frequency from the
initial conditions. $\bm{a}$ in \eqr{fpo} is defined as
\begin{equation}\label{eq:a}
  \frac{1}{2}\bm{a} = \Gamma \nabla_v h(\bm{x},\bm{v}),
\end{equation}
where $\Gamma = 4\pi \Lambda (Ze)^4/m^2$ ($Z$ is assumed 1 here) and
$\Lambda$ is the Coulomb logarithm. Since $h$ is defined as
$\nabla^2h=-f$, it can be calculated analytically for the Maxwellian
distribution $f$. Without loss of generality, we can assume a zero
bulk velocity to get,
\begin{equation}\label{eq:maxwellH}
  h_M(\bm{x},\bm{v}) =
  \frac{n(\bm{x})}{v}\mathrm{erf}\left(\frac{v}{\sqrt{2}v_{th}(\bm{x})}\right),
\end{equation}
where $v=\sqrt{v_x^2+v_y^2+v_z^2}$.  Taking the velocity gradient
from \eqr{a} leads to
\begin{equation}\label{eq:diff_maxwellH}
  \nabla_v h_M(\bm{x},\bm{v}) =
  \left[\frac{\sqrt{2}}{\sqrt{\pi}}\frac{n}{v_{th}}
    \frac{\exp\left(-\frac{v^2}{2v_{th}^2}\right)}{v^2} - n
    \frac{\mathrm{erf}\left(\frac{v}{\sqrt{2}v_{th}}\right)}{v^3}\right].
\end{equation}
Comparing the forms of \eqr{colld} and \eqr{fpo} leads to
the effective collision frequency,
\begin{equation}\label{eq:freq}
  \nu_{ef}(\bm{v}) = -\Gamma\left[\frac{\sqrt{2}}{\sqrt{\pi}}\frac{n}{v_{th}}
    \frac{\exp\left(-\frac{v^2}{2v_{th}^2}\right)}{v^2} - n
    \frac{\mathrm{erf}\left(\frac{v}{\sqrt{2}v_{th}}\right)}{v^3}\right].
\end{equation}
$\nu_{ef}(\bm{v})$ is the effective collision frequency between a test
particle at $\bm{v}$ and a non-drifting Maxwellian with number density
$n$ and thermal velocity $v_{th}$. Since the goal of this
approximation is to estimate the effect of collisions between the two
jets, i.e., the collision frequency between a particle in one jet and
particles in the other jet, we can set $|\bm{v}|=\Delta u = 2u$
(internally, the jets are collisional).

Interestingly, \eqr{freq} gives comparable results to the Spitzer
formula of the collision frequency for the slower jets
(\SI{15}{km/s}).  However, the collision frequency decreases
approximately by an order of magnitude for the faster jets
(\SI{50}{km/s}) where the Spitzer formula would not apply.

\eqr{freq} assumes a Maxwellian distribution of particles and is only
valid before the jets start to merge.  Therefore, the full FPO is
required to study the details of the shock-forming process.  Still,
calculating the effective collision frequency, $\nu_{eff}$, and using
it in \eqr{colld} is sufficient to answer the question of
\textit{whether} the jets are going to form a shock or just
interpenetrate for a range of parameter scans across density,
temperature, and velocities.  Therefore, it can be used to quickly
guide an experimental setup and perform broad parameter
studies. $\nu_{eff}$ is used exclusively in the rest of this work. In
follow-up work, we will complement this approximation with the full
FPO simulations of selected cases.

\section{Scaling study}\label{sec:scaling}

\begin{figure}
  \centering
  \includegraphics[width=\linewidth]{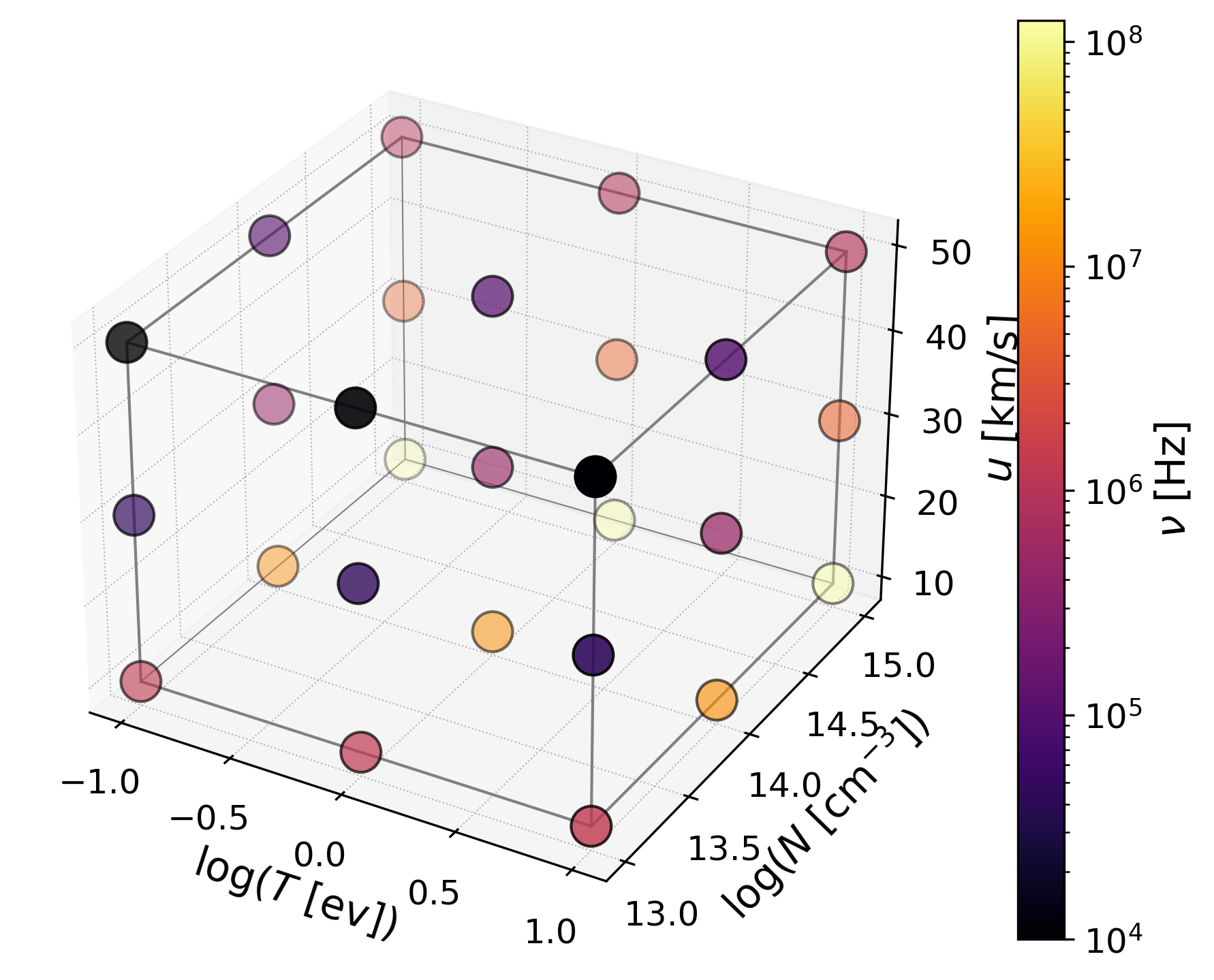}
  \caption{Effective collision frequencies from \eqr{freq} calculated
    for the initial parameter space defined by temperature, density,
    and jet bulk velocity.  Note that the scales of the density,
    temperature, and collision frequency are logarithmic.}
    \label{fig:space}
\end{figure}

Using the frequency \eqr{freq}, we can set up a wide variety of
relatively fast numerical simulations. Upon merging, the variations in
the temperature, density, and bulk velocity ($|\bm{v}|=\Delta u = 2u$)
result in either an interpenetration of the jets or an increase in the
density. Figure\thinspace\ref{fig:space} shows a 3D sample of a
parameter space where the effective collision frequency is
color-coded. This illustrates that the collision frequency drops by
over an order of magnitude between the 15 and \SI{50}{km/s} jets.

\begin{figure}
  \centering
  \includegraphics[width=.9\linewidth]{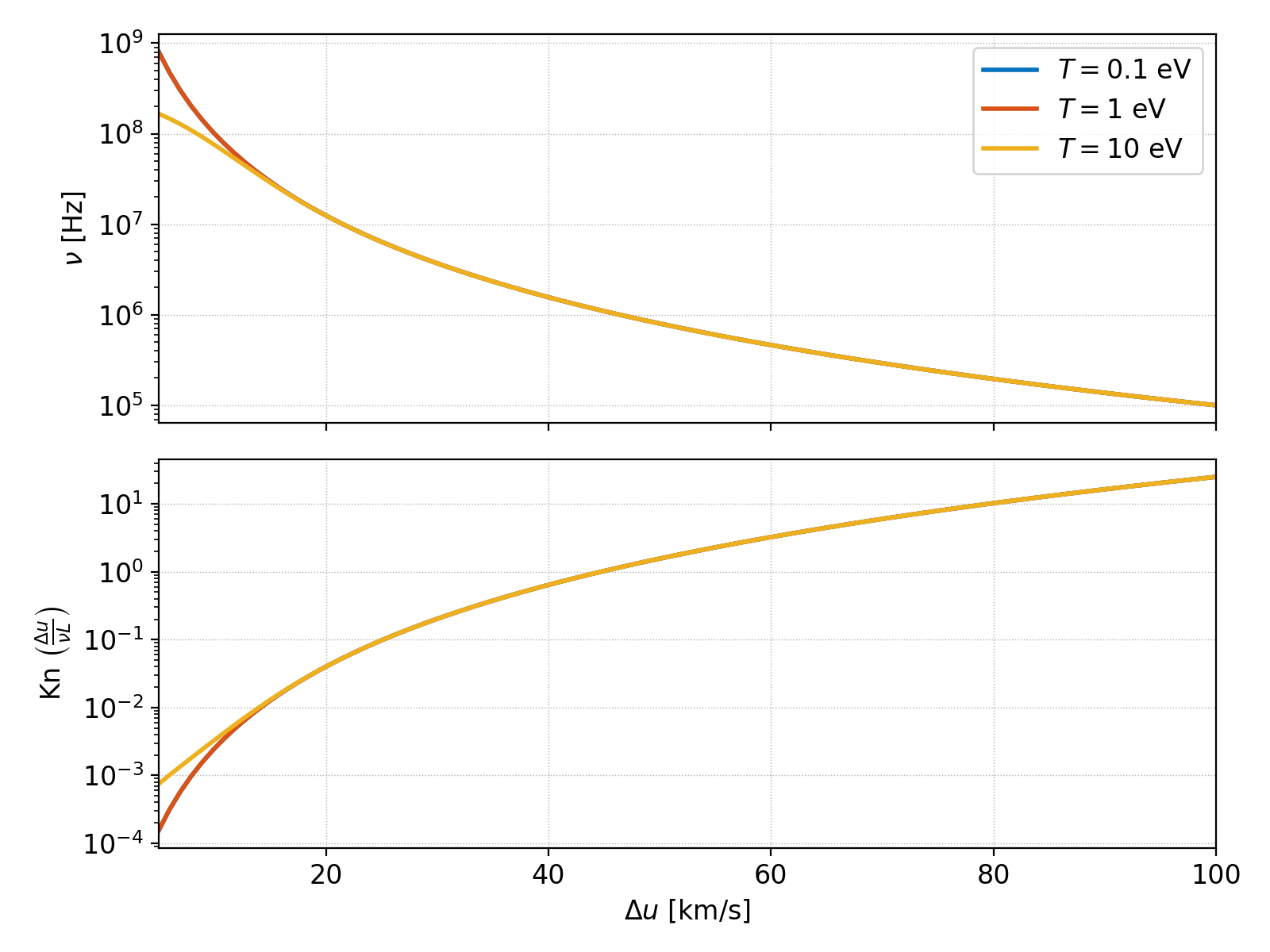}
  \caption{Ion-ion collision frequencies (top panel) from \eqr{freq}
    for density of \SI{1e14}{cm^{-3}} and three temperatures, 0.1, 1,
    and \SI{10}{eV} based on the \textit{relative} velocity (for
    symmetric jets $\Delta u = 2u$). All three temperatures give
    identical results for relative velocities higher than
    \SI{20}{km/s}.Bottom panel shows corresponding Knudsen number
    where the mean free path is calculated as $\Delta u / \nu$ and $L$
    is taken as the simulation domain size of \SI{4}{cm}.}
  \label{fig:coll}
\end{figure}

A particular feature of \fgr{space} is the independence of the
effective collision frequency on the temperature (for these jet
velocities). This is consistent with the low thermal velocity relative
to the bulk velocity, i.e., the argon ions behave like cold
beams. Figure\thinspace\ref{fig:coll} shows the dependence of the
collision frequency from \eqr{freq} on the \textit{relative} velocity
(for symmetric jets $\Delta u = 2u$). The three studied temperatures
(0.1, 1, and \SI{10}{eV}) give identical results for relative
velocities higher than \SI{20}{km/s}. The lower two temperatures
effectively match across the whole range of relative velocities.

\begin{figure}
  \centering
  \includegraphics[width=.9\linewidth]{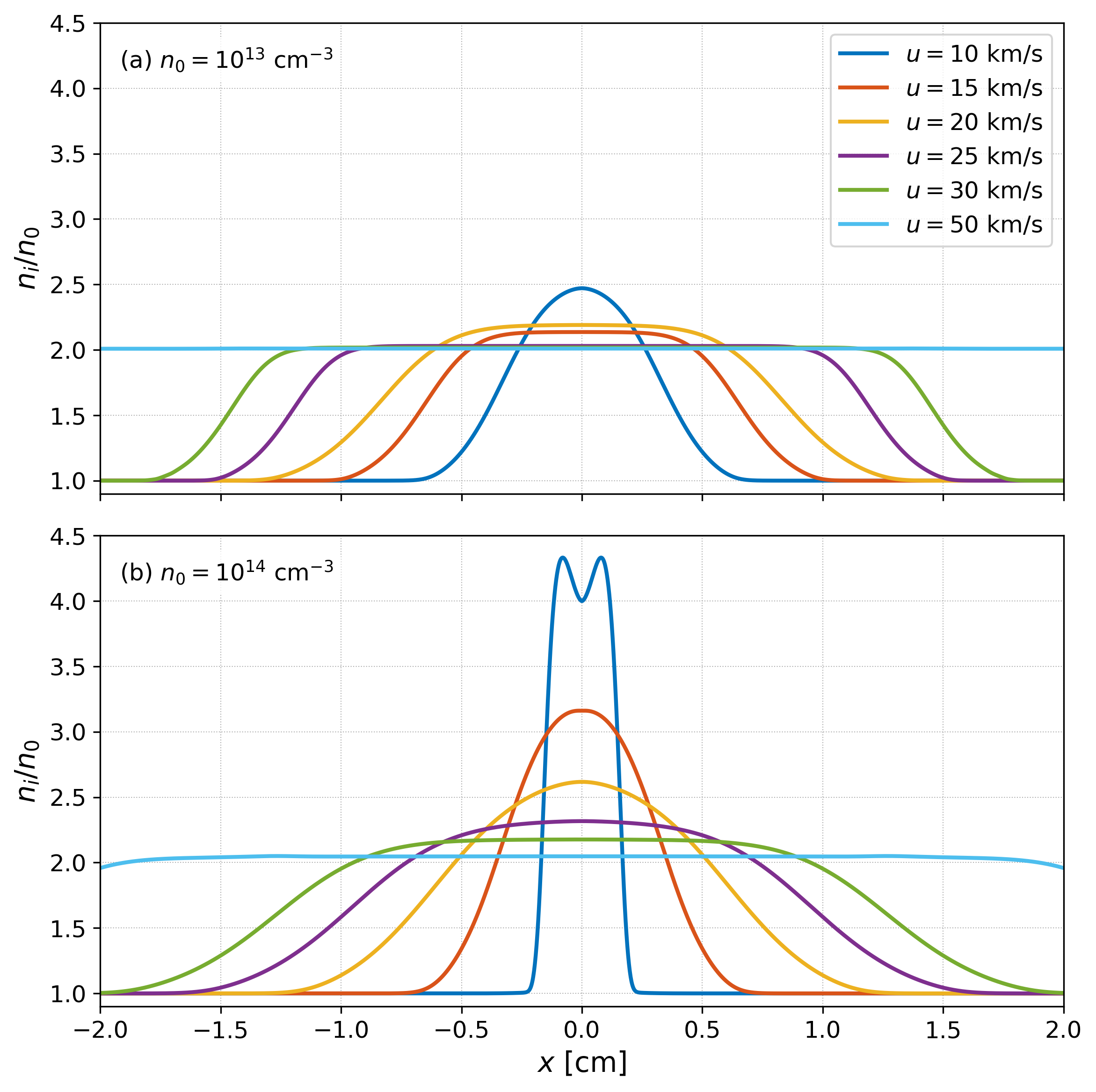}
  \caption{Comparison of densities for several jet velocities after
    \SI{0.5}{\micro s}. Two initial densities, \SI{1e13}{cm^{-3}} and
    \SI{1e14}{cm^{-3}}, are shown. Only lower velocities for the lower
    density case are merging while the higher velocity jets are simply
    streaming through each other.}
  \label{fig:scan}
\end{figure}

With the temperature having a negligible impact on the collision
frequency in the regimes considered, several additional simulations
for different jet velocities are performed to help find the transition
point between interpenetration and shock
formation. Figure\thinspace\ref{fig:scan} and \fgr{scan2} show two
initial densities, \SI{1e13}{cm^{-3}} and \SI{1e14}{cm^{-3}} for the
temperature of \SI{1}{eV}.

Figure\thinspace\ref{fig:scan} captures the number densities at
\SI{0.5}{\micro s} after the contact. Note that in a fluid simulation
with Euler equations, the density would quickly jump by a factor of
four and form sharp gradients; however, this is not the case here due
to the finite collision frequency. The lower density, \fgr{scan}a,
results in lower collision frequency and the jets mostly
interpenetrate. Only the lower jet velocities have density
accumulation in the middle, where the density goes above the
superposition of the two jets. This density is going to increase for
as long as there is a flow of plasma but the feature is going to be
diffused. In the case of the higher density, \fgr{scan}b, there is a
transition velocity below which there is significant accumulation in
the center and sharp features arise.  Similar to the lower density
case, the number density will increase for as long as the source of
plasma is present. As it was mentioned above, the goal of this method
is to capture the transition between the shock and interpenetrating
regime, not to study subsequent details of the shock
formation. Therefore, the results for cases where density
significantly increases should be seen as qualitative.

In addition to densities in \fgr{scan}, \fgr{scan2} provides line-outs
of the ion distribution function in the middle of the domain.  The
line colors are consistent with those in \fgr{scan}.  The top panel
shows that the only case with density growth (the \SI{10}{km/s} case)
is also the only one where the ion jets are merging. The jets remain
distinctly separate in all the other cases.  The distribution
functions in the higher density case are clearly merging to form a
central shocked region for most of the jet velocities considered.

\begin{figure}
  \centering
  \includegraphics[width=.9\linewidth]{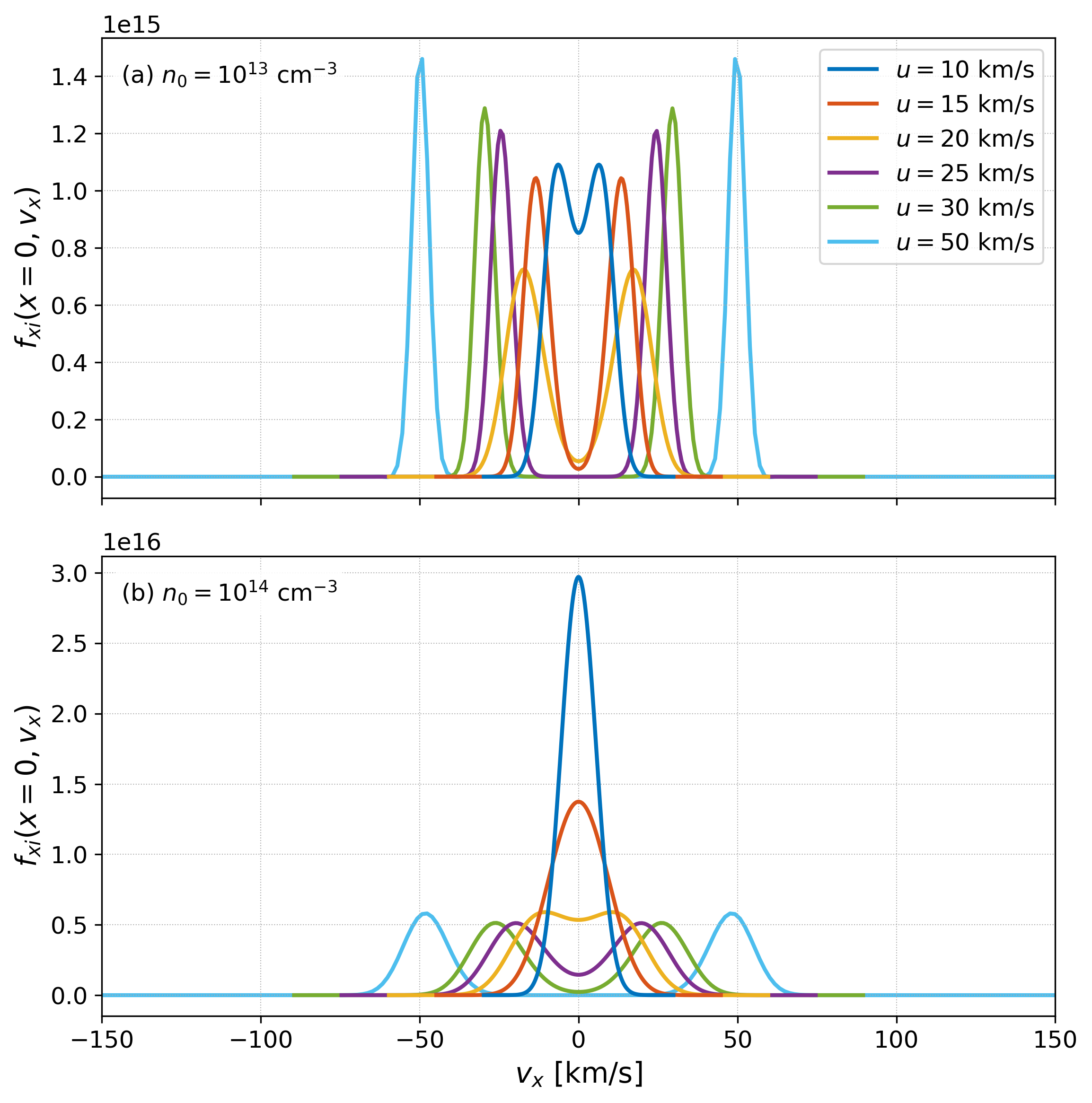}
  \caption{Comparison of distribution function lineouts at $x=0$ for
    several jet velocities after \SI{0.5}{\micro s}. Two initial
    densities, \SI{1e13}{cm^{-3}} (top) and \SI{1e14}{cm^{-3}}
    (bottom), are shown. These plots show a clear difference between
    the cases where jets merge and where the jets interpenetrate. Note
    that for the lower density, the collision frequency is lower and
    even the slowest tested jet does not fully merge in time, still
    presenting double peaks.  In the higher-density case, the two
    slowest cases are fully merged.}
  \label{fig:scan2}
\end{figure}

\section{Comparison to experimental data}\label{sec:plx}

\begin{figure}
  \centering
  \includegraphics[width=.9\linewidth]{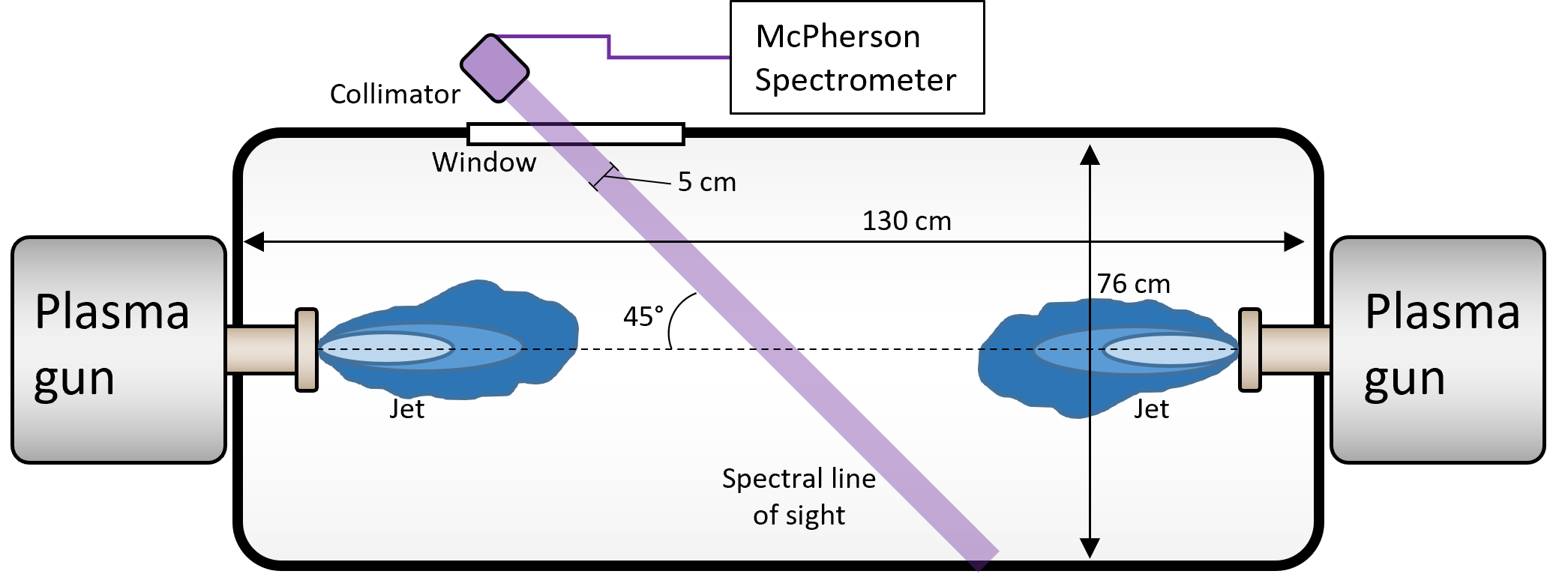}
  \caption{A diagram of the experimental of the jet merging chamber
    that uses plasma guns from PLX. The purple highlighted region
    indicates the spectral line of sight approximately \SI{5}{cm} wide
    through which the Doppler shift measurements are acquired.}
  \label{fig:experimentsetup}
\end{figure}

\begin{figure}
  \centering
  \includegraphics[width=.9\linewidth]{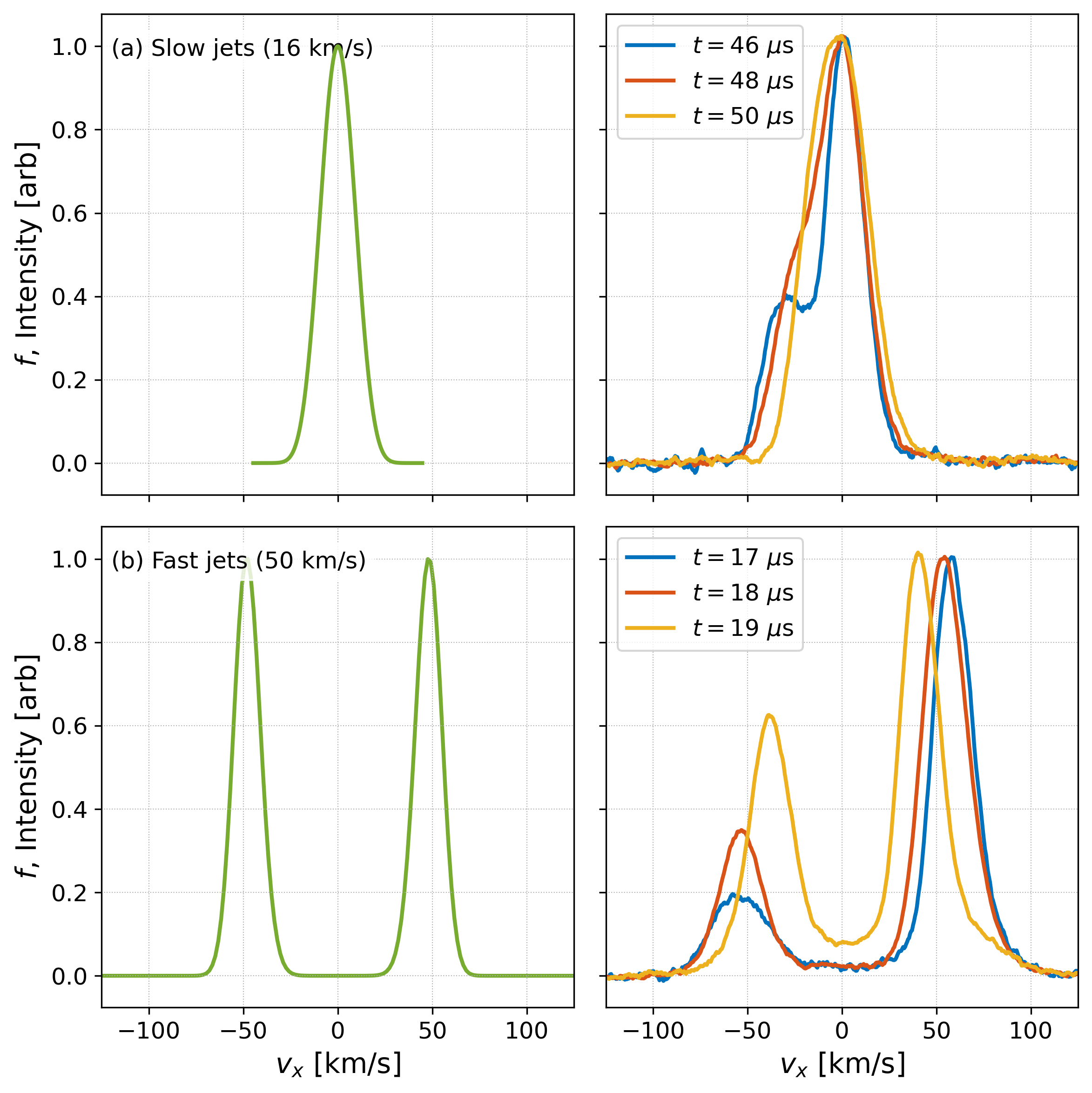}
  \caption{Doppler shift measurements of the ArII \SI{434.8}{nm}
    emission from PLX (right column) which directly correlate to the
    simulation distribution function as seen in the left column and
    \fgr{scan2}(b).} These results agree with the model predictions
  where the slow jets merge while the faster jets interpenetrate. The
  time in this figure is measured from the shot and not from the point
  of contact as is the case for the simulation results.
  \label{fig:experiment}
\end{figure}

The simulations in the previous section demonstrate the potential of
the hybrid model with a reduced number of fully resolved velocity
dimensions to predict different jet-merging regimes. In order to
validate our results, we look at the preliminary PLX data from shots
with two jets with normal incidence, which are the closest to the
simulation setup. The experimental chamber is cylindrical, \SI{76}{cm}
in diameter and \SI{130}{cm} long, and uses plasma guns from PLX to
recreate similar conditions, but is itself separate from the PLX
chamber. Figure\thinspace\ref{fig:experimentsetup} shows a diagram of
this chamber along with the spectral line of sight from which Doppler
shift measurements are obtained. At each end is mounted one of the
plasma guns which launch the jets. Gas feed pressure and discharge
voltage of the guns can be adjusted to vary the jet speed and explore
the breaking point between merging and interpenetrating
jets. Figure\thinspace\ref{fig:experiment} shows intensity data from
the Doppler shift measurements, where the wavelength has been
converted to velocity. The spectral line of sight is 45 degrees with
respect to the jet axis, which is accounted for in converting from
Doppler wavelength shift to the axial ion velocity distribution in the
figure. There we clearly see the slower jets merging (top) and faster
jets retaining two peaks (bottom); they do not merge even later in
time. Note that the times in this figure are measured from the shot
and not from the point of contact as is the case for the simulation
results. The intensity asymmetry of the distribution functions is most
likely caused by the line-of-sight of the diagnostics being under an
angle with respect to the jets, with the farther jet contributing a
lesser overall intensity.

\section{Closing remarks and Summary}

\begin{figure}
  \centering
  \includegraphics[width=.9\linewidth]{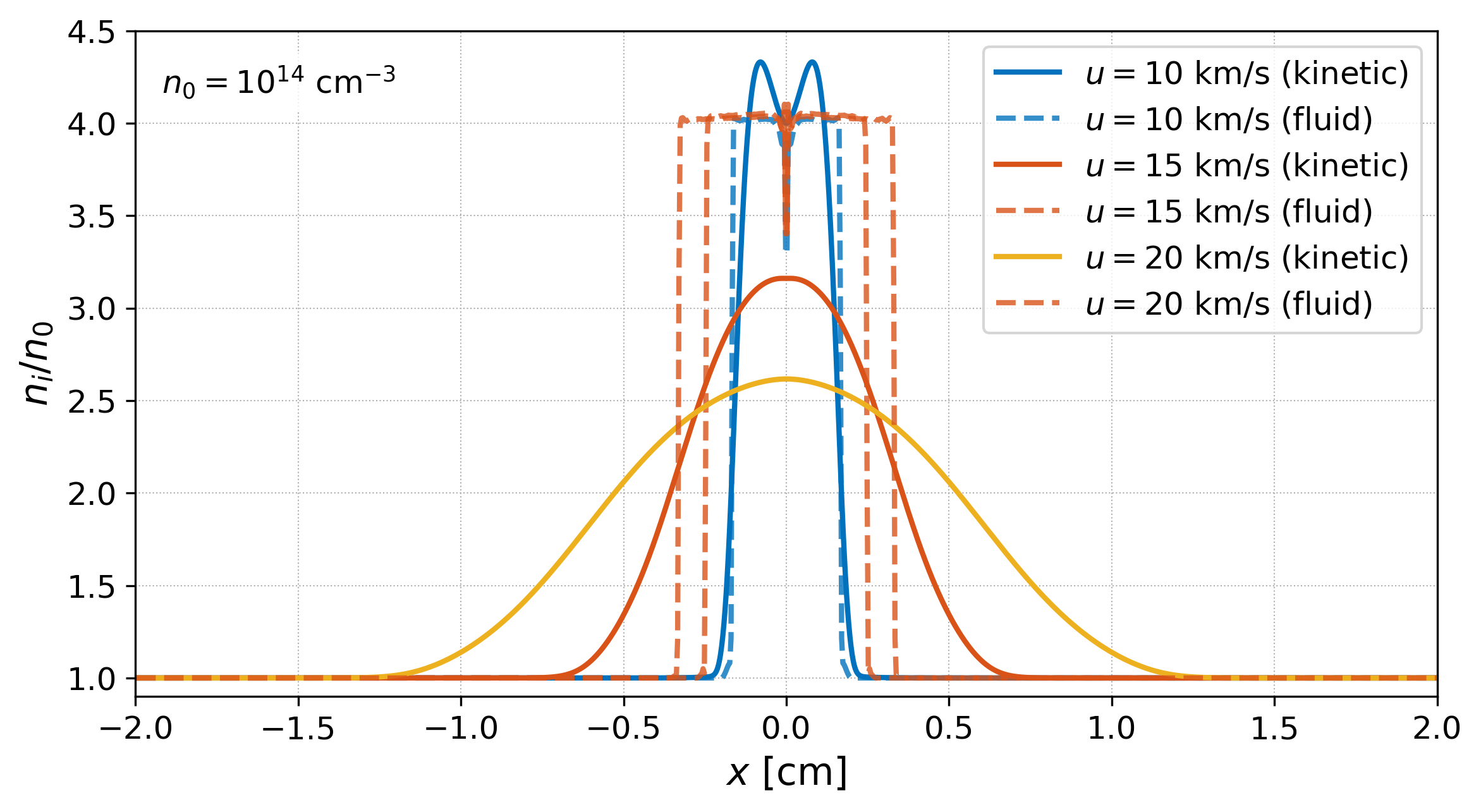}
  \caption{The exact density profile as seen in \fgr{scan} with an
    addition of densities from the \gke{} five-moment fluid
    model. Note that even if the fluid model is started with exactly
    the same parameters, it would always result in a shock and sharp
    gradient. The kinetic results with finite collision frequency are
    significantly more diffused.}
  \label{fig:fluid}
\end{figure}

It is important to note that when these cases are simulated using a
five-moment two-fluid model,\cite{Hakim2006high,
  srinivasan2011analytical, shumlak2011advanced} the merging of the
jets will always produce a shock with sharp density
gradients. Figure\thinspace\ref{fig:fluid} shows the three cases with
higher density from \fgr{scan}b along with results from the
five-moment two-fluid model implemented in \gke{}. While the position
of the gradients does match for the slowest case, where the collision
frequency is relatively high, the other cases produce very different
results.  As a result, fluid models cannot be used to accurately
predict shocked versus shock-mitigated (or interpenetrating) jet
merging.

To summarize, the transition between collisional and collisionless jet
merging is vital to understand in a variety of plasma systems.  To
accurately model these phenomena one requires not only a kinetic
method but a kinetic method that accurately describes all of the
degrees of freedom of the plasma. We have demonstrated such a model
with our novel parallel-kinetic-perpendicular-moments model that
decomposes the collisionless dynamics parallel and perpendicular to
the jet merging process.  We find that slow jets merge and shock while
faster jets are sufficiently collisionless to simply interpenetrate.
These results are in agreement with PLX measurements.

We emphasize that the kinetic model presented in this
  paper is unique not only in its capability of simulating the
  transition between collisional and collisionless jets, but in the
  quality of the distribution function data the method produces from
  the grid-based approach employed. Thus, the grid-based method
  employed here not only contains the key kinetic physics required to
  model these systems but also eliminates the counting noise typical
  of other numerical approaches such as the particle-in-cell method
  which can be problematic for analysis of the distribution function
  and overall solution quality.\cite{Juno2020noise} We anticipate this
  approach to be of high utility for future jet merging studies which
  include geometric effects such as the angle of jet merging.

\section{Getting \gke{} and reproducing the results}

To allow interested readers to reproduce our results, full
installation instructions for \gke{} are provided on the \gke{}
website (\url{http://gkeyll.readthedocs.io}). The code can be
installed on Unix-like operating systems (including Mac OS and Windows
using the Windows Subsystem for Linux) by building the code via
sources.  Input files for the simulations presented here are available
in the following GitHub repository,
\url{https://github.com/ammarhakim/gkyl-paper-inp}.

\section{Acknowledgment}

This work was supported by the Department of Energy ARPA-E BETHE
program under award number DE-AR0001263.

The authors acknowledge Advanced Research Computing at Virginia Tech
for providing computational resources and technical support that have
contributed to the results reported in this paper. URL:
\url{https://arc.vt.edu}

\bibliography{reference}

\end{document}